# Fundamental Analysis in China: An Empirical Study of the Relationship between Financial Ratios and Stock Prices

Lijuan Ma[1], Marcel Ausloos[1], Christophe Schinckus[2], H. L. Felicia Chong[3]

[1]School of Business, University of Leicester, Leicester, UK
[2]Department of Economics and Finance, RMIT University Vietnam, Ho Chi Minh, Vietnam
[3]Institute of Graduate Studies, University of Malaya, Kuala Lumpur, Malaysia

Abstract

The informational context is regularly questioned in a transitional economic regime like the one implemented in China or Vietnam. This article investigates this issue and the predictive power of fundamental analysis in such context and more precisely in a Chinese context with an analysis of 3 different industries (media, power, and steel). Through 3 different kinds of correlation, we examine 25 financial determinants for 60 Chinese listed companies between 2011 and 2015. Our results show that fundamental analysis can effectively be used as an investment tool in transitional economic context. Contrasting with the EMH for which the accounting information is instantaneously integrated into the financial information (stock prices), our study suggests that these two levels of information are not synchronized in China opening therefore a door for a fundamental analysis based prediction. Furthermore, our results also indicate that accounting information illustrates quite well the economic reality since financial reports in each industry can disclose a part of stock value information in line with the economic situation of the industry under consideration.

1. Introduction

Valuation of assets is a key issue in finance and a plethora of predictive tools can be considered to be very useful for financial management. In mature markets, Fama [1] [2] explained that the security's price given by the market can be considered as its intrinsic value. This famous claim is now well known as "the efficient market hypothesis" (EMH) according to which the market provides the assets' price by integrating all kind (public and/or private) of information related to them; for a recent discussion, see Jovanovic and Schinckus [3]. Counterarguments have been suggested, a famous one being claimed by Grosmann and Stiglitz [4] who explained that investors have no interest gathering information unless their efforts are compensated with higher return. Authors wondered how prices can reflect the intrinsic value if no information is gathered [4] [5]. Beyond this paradox, the environment (or microstructure) defining the market rules does not necessarily offer the institutional conditions favouring the observation of EMH.

China is an example of such situation. Indeed, although China has a high economic growth, the country does not offer a strictly marked based environment. Precisely, China is said to have a transitional regime in which markets play an important role but the State remains the major economic actor. In other words, this particular institutional context makes the EMH irrelevant in China, opening therefore some doors for prediction of market prices. Technical analysis and fundamental analysis are two well known predictive methods. While the former is associated with the identification (and repetition) of visual patterns in the financial dynamics of prices [6], the latter rather refers to a method evaluating securities through the analysis of financial statements.

Our paper deals with fundamental analysis, considering that, in a transitional economic context, earnings, dividends, and other financial ratios coming from financial statements can be useful in the valuation and prediction of securities' prices. Although a large literature exists on the importance of fundamental analysis, few studies have been made in a transitional context. However, this issue is important for two reasons, on the one hand it characterizes several countries such as China or Vietnam; and on the other hand, it directly questions the quality of accounting information in countries led by such regimes. In other words, this article contributes to the existing literature on two aspects: the predictive power of financial ratios/statements and the quality of the information provided by these financial statements in a transitional economic configuration, and more precisely China. This issue is very important in China where informational context is regularly questioned [7].

This paper investigates these aspects by focusing on the application of the fundamental analysis in a specific industrial environment defined by 3 major sectors (media, power and steel industries) in China between 2011-2015. This specific time window is justified by the existence of the Chinese stock markets turbulence in 2015-2016. Precisely, this financial crisis makes quite difficult all kinds of study in terms of fundamental analysis.

The section overviews the major works dealing with the predictive power of financial ratios coming from financial statements in an emerging (and sometimes developed) country context, while the third section presents our data for the 25 selected determinants. In the fourth section, we discuss our findings while the last section ends this study with some conclusive remarks.

2. Literature Review

Broadly speaking, a fundamental analysis can be done at three levels: micro, industrial or macroeconomic. The macroeconomic analysis refers to all indicators related to the economic environment in which the companies are operating. Key actors such as central banks are important since they use rediscount policy, open market operation, and legal reserve ratios as means to adjust the supply of currency in order to control the liquidity and maintain a steady economic development. Adjustment of all benchmark rates can affect a stock price. For example, the investment value of a security may decrease during an increasing period, bringing about stock price dropping; in the contrary situation, the result shows an opposite direction [8]. Numerous empirical studies exist on the potential influence of inflation/level of GDP on the evolution of the assets' price (see Weale and Wieladek [9] for a recent analysis of these aspects).

At the micro level, the relationship between financial ratios and stock prices has been extensively studied in the literature since Ball and Brown [10]. They adopted the method of sign test to study stock prices over 12 months prior to the release of accounting information and for 6 months after the release, by using 216 companies selected from the New York Stock Exchange [10]. Their results show a positive correlation between the trend of financial information changes and the changes in the stocks' price. In the same year, Beaver [11] studied the change of the day trading volume due to the disclosure of earning information for 506 groups. He found that the volume of information may boost the share price [11]. Later, Beaver, Clarke, and Wright [12] divided 276 public firms into 25 portfolios to research whether a relation exists between unsystematic returns and the magnitude of earnings forecast errors. As a result, it was found that the Spearman rank correlation coefficient of the percentage forecast errors is equal to 0.75 implying a significant statistical result. Precisely, there exists a positive reciprocal relation between these two variables. More recently, the financial ratios and share prices relation was analyzed (over a time span between 2003/1 and 2011/1 for European and S & P500 stocks) in order to reflect the performance and value of companies in financial ratios before the selection of portfolios' structure [13] [14]. O'Hara et al. [15] believe that both current profits as well as the tendency of gains may work in deciding stock prices. According to these authors, however, the price fluctuation might not synchronize immediately. Therefore, there is a room for an influence of the price of securities on expected profits.

Other researchers studied the correlation between stocks' prices and accounting indicators. Such studies indicate different correlations and sensitivities depending on countries and industries. For example, Aono and Iwaisako [16] demonstrated that the influence of price earnings ratios is quite strong in USA, while this ratio exerts a weak influence on stock prices in Japan. Moreover, price earnings ratios differ from other ratios playing a crucial role in stock price, especially for smaller corporations [16]. On the same topic, Pech, Noguera and White [17] or Lewellen [18] find a close relationship between stock prices and ratios, such as dividend yield, earning per share, and book to market value of equity. Jiang and Lee [19] indicated that several value models, such as the dividend discount model, the earning discount model, and the residual income model, use financial ratios as a theoretical basis to predict market trends, but challenged that financial ratios, since being based on fixed rates during the accounting period, are difficult to be correlated with stock returns. Regarding the emerging financial markets, Martani, Mulyono, and Khairurizka [20] showed that stock market evolution can be predicted using activity ratio: profitability, liquidity, and also leverage, market ratio, size and cash flows). These results were also consistent with previous studies like Höbart [21] and Bakti and Sumaedi [22]. Petcharabul and Romprasert [23], for instance, showed the existence of a clear relationship between 5 financial ratios and stock returns in technology industry of the Stock Exchange in Thailand (SET) from year 1997 to 2011 for 15 years (quarterly data from financial statement). Bakti and Sumaedi [22] looked at the effect of 7 financial ratios on stock returns for ISO 9001 Certified Companies' traded on the Indonesian Stock Market between 2007 and 2009, but found no impact on stock returns.

Concerning the Chinese market, it appears to be very specific in this context. Indeed, several authors [24] [25] [26] [27] showed that the EMH cannot be assumed in the same way than in the Western markets, in particular concerning the synchronicity of stock price movements. Based on regression results, among dozens of financial variables which were tested, Zhang [28] concluded that the year to year revenue growth is found to be the most significant variable for predicting stock prices. This paper aims at studying the predictive power of financial ratios in China through a statistical analysis of the relationship between stock prices and financial ratios for the 3 major Chinese sectors as detailed in the following sections.

3. Methodology and Data

The panel data rely on Ashare stocks traded in the Shanghai and Shenzhen stock exchanges between 2011 and 2015. This sample is divided into three groups according to industry (media, power and steel). A set of 20 companies, with public, accurate and financial information as well as authentic, accurate and complete information disclosure was selected for each group. For firms with incomplete or confusing data, they have been removed from the sample. Regarding the missing data in the stock markets, we replaced them by the average of the three previous periods in line with the existing empirical studies on the topic [28]. We finally have 55 companies in the media industry; 95 in the power industry and 88 companies for the steel industry. Regarding our accounting determinants, they can be grouped into five categories: (1) profitability ratios, (2) liquidity ratios, (3) operating capacity ratios, (4) development ability ratios and (5) solvency & risk ratios, as detailed in the following subsections.

3.1. Selection of Dependent Variables

Stock prices are regarded as the dependent variables of the model. In the same vein, the closing price of the next trading day, after the disclosure of the annual report, is treated as a dependent variable. This value seems to us the best one reflecting how much the share price is immediately affected by the annual financial report. 3.2. Selection of Independent Variables The "independent variables" are the accounting ratios; they can be calculated from the annual report of such listed companies.

(1) Profitability ratios

This paper selects four ratios to reveal the profitability of sample firms [29]: (1) gross profit rate, (2) net profit rate, (3) return on assets (ROA), and (4) return on equity (ROE). The more benefit a company provides for its shareholders, the more popular it will become in the stock market [29]. In this paper, this ability is measured by (5) earnings per share (EPS) and (6) book value per share (BVPS) (Table 1).

(2) Liquidity determinants

Large cash flows indicate that the company is able to repay debt and/or to invest and has the capability of suitable change in the economic environment [15]. This paper uses (7) cash ratio, (8) cash maturing debt ratio, (9) debt coverage ratio, and (10) net cash flow per share to reflect a company indicator of generated cash flow (Table 2).

(3) Operating capacity ratios

A firm operating capacity can be described by four ratios: (11) turnover of account receivable, (12) inventory turnover, (13) current assets turnover, as well as (14) the total assets turnover (Table 3).

(4) Development ability ratios

Companies with well operating performance may achieve improvement faster than others, thereby attracting more participants to invest. A company development ability can be typically reflected by 4 ratios: (15) rate of capital accumulation, (16) the growth rate of EPS, (17) the growth rate of ROE, and (18) its net profit growth rate (Table 4).

(5) Solvency & risk

Solvency is an essential index when considering the financial condition of a listed company. Without a solid solvency basis, a company may confront some risk of bankruptcy and not being capable of paying its debt on time. Solvency ratios can be divided into short term solvency indicators and longterm solvency indicators. This article considers (19) the current ratio and (20) the quick ratio in order to measure short term solvency. The longterm ratios include: (21) the debt-to-assets rate and (22) the equity ratio. Related to the solvency, we can mention the risk of investment is considered to be one of the most significant factors that influence investors' decisions.

When investing in a company, leverage coefficients reflect the risk sensitivity during corporate operation. The risk level of companies is usually measured through (23) the financial leverage, (24) the operating leverage, and (25) the degree of total leverage (Table 5).

All these indicators deduced from firms' economic activity (collected from their financial statements) help us to investigate the potential predictive power of a fundamental analysis in the Chinese context. The empirical analysis follows in the fourth section hereafter.

4. Empirical Analysis

The descriptive statistical analysis and the correlation analysis have followed standard procedures. We compute the characteristics of the distributions of ratio values for each year (minimum, maximum, mean, and standard deviations) for each industrial sector, respectively. In particular the correlations between the closing value of the day following the annual report release and the ratios found in (or estimated) from such reports have been analyzed through the Pearson r coefficient, on one hand, and through the rank-rank correlation Kendall τ and Spearman ρ method, on the other hand; all 3 coefficients are useful for pointing to disparities and similitudes [30].

The 3 following subsections summarize the major results of our analysis for each industry, i.e., media, power, and steel, respectively. The comments are deduced from a global analysis, but only the most significant correlations (distinguishing 0.01% and 0.05% levels) are reported in the Tables found in the Annexes (for saving space and providing a better emphasis on the correlation coefficient values). It is a useful advice to demand the reader to observe the sign of each coefficient, beside their value and significant level.

4.1. Empirical Analysis of the Media Industry

The correlation analysis for the media industry can be found in Annex 1. Few cases turn out to be significant. In 2011, merely two factors, turnover of current asset and turnover of total assets, from a total of 25 ratios of annual reports of public media firms, passed the Pearson test. The relationship between these two ratios and the stock price presents a relatively significant correlation at a 0.05 level; the correlation coefficients equal to 0.809 and 0.754, respectively. The Kendall test and the Spearman correlation test for the 25 ratios indicate that only the degree of operating leverage, total sale change and comprehensive leverage coefficient pass these tests, at the 0.05 significant level.

In the correlation test, only the current asset turnover displays a relatively strong positive relation with price, for a correlation coefficient = 0.932 at the significant 0.01 level. There are six ratios showing a weak influence on the share price, among which the cash ratio presents the strongest negative relevance to the price. The cash ratio, turnover of current asset, current ratio, and quick ratio are found to have a significant correlation, both according to the Kendall and the Spearman correlation tests. The debt-to-asset and the equity ratio also present some impacts to the investment value.

Remarkably, 24 factors (all 25 except the rate of capital accumulation) do not have a relation with stock prices. The correlation coefficient or the rate of capital accumulation in the Pearson test is 0.809, the significant level of which is 0.015. About the Spearman correlation test, the inventory turnover is the only ratio having an influence on the price in 2013. It seems that one can reasonably conclude that none of these ratios can be reflected on the share price value of the public media companies, in 2013.

Among the four indicators related to ability, investors need to pay more attention on changes of the EPS and changes in the net profit rate. For the Pearson test, the correlation coefficient of change of the EPS and price achieves a significant 0.625 value, at the 0.007 level, while the coefficient of the change in the net profit rate as well as in the investment value reaches 0.644, at a significant 0.005 level. The rates of capital accumulation are found not only to have a correlation with the price. In addition to these, BVPS and the degree of operating leverage also show a significant influence on the share price at the 0.03 level. However, other figures do not significantly pass the correlation study of the Pearson test, nor the rank correlation Kendall test or the Spearman correlation test. They are thus not reported in Tables A1-A4, in Annex 1.

4.2. Empirical Analysis of the Power Industry

The correlation analysis for this industry can be found in Annex 2, Tables A5-A9. For 2011, a positive correlation of financial ratios and stock price is observed. Merely 7 factors, that is, ROA, ROE, EPS, net cash flow per share, inventory turnover ratio, turnover of current asset as well as the turnover of total capital, passed the Pearson test.

The correlation results from the relationship between ROE, EPS as well as net cash flow per share and share price respectively achieve the significant level of 0.01. The strongest correlation is the relevance of the EPS and investment value, having a correlation coefficient equal to 0.839. The other four factors show a weaker relationship with stock prices. Only the net cash flow per share, the turnover of inventory and the turnover of current assets display a negative relation with the share price. The "first rank" is for the correlation between the net cash flow per share and the share price, at a significant level equal to 0.007: the correlation coefficient is equal to −0.615. Based on the Kendall test and the Spearman test, several (11) factors indicate a significant correlation with the stock price. There are ROA, ROE, EPS, net operating cash flow, total debt, inventory turnover ratio, turnover ratio of total assets, change of EPS, current ratio, quick ratio, and equity ratio. The inventory turnover ratio, turnover ratio of total assets, and current ratio have a stronger impact on the share price than the other 8 factors, at the significant level of 0.01; especially, the correlation coefficient of the turnover ratio of total assets with respect to the price equal 0.502 and 0.707, using the Kendall test and the Spearman correlation test, respectively.

Profitability, benefit ability to shareholders, cash flow index and solvency are regarded as important factors in relation to stock prices; they passed the Pearson test, and also the Kendall and the Spearman test. Regarding the Pearson test, ROA, ROE, EPS and the net cash flow per share indicate a close relation with prices, at the significant level of 0.01, while cash flow ratio, cash maturity debt ratio and current ratio should be less relevant for investors' strategy than the former four.

The correlation coefficient of EPS holds the "first rank" among the factors when considering the same significant level for all cases. The operating capability and development ability have an influence on the share prices, when using the Kendall or the Spearman correlation test. Other factors do not present any significant correlation with the stock prices in the power industry sector.

The trends in the evolution of the stock prices in 2013 exhibits a strong relation with the financial ratios obtained from annual reports of power companies. However, three indicators are found to be significant at the 0.01 level for the Pearson test: the Cash ratio, Current ratio and Quick ratio. The Current ratio presents the most significant relevance to the share price at the significant 0.01 level since the correlation coefficient is 0.744. Apart from the Current ratio, the Cash ratio and the Quick ratio, Net operating cash flow and Total debt, the Kendall test and Spearman correlation test shows a high relevance for the stock prices. The Total debt shows the strongest negative relationship, thereby not suggesting an investment.

The cash flow index and the operational capability present a significant correlation to stock price, at the 0.01 level, based on the Pearson test. However, merely the turnover ratio of total assets shows a positive relationship with prices, reaching a correlation coefficient value equal to 0.692. The cash maturity debt ratio, turnover ratio of total assets, current ratio, quick ratio, and EBIT have a weak influence on the stock price.

Differently from the result of the Pearson test, the current ratio can be regarded as an important factor when considering investment on power companies by using the Kendall or the Spearman test. Apart from these, the net cash flow per share passed the

Spearman test, showing a reasonable relationship to investment value, at the significant level of 0.041 with a 0.332 correlation coefficient.

Our observations suggest that 11 factors, that are, the Cash ratio, Cash maturity debt ratio, Net operating cash flow indicator, Total debt ratio, Turnover ratio of total assets, Rate of capital accumulation, Growth rate of net profit, Current ratio, Quick rate, Degree of operating leverage, and Comprehensive leverage coefficient passed the 3 correlation tests. However, only the Turnover ratio of total assets, the Current ratio and the Comprehensive leverage coefficient are found to reach the 0.01 significant level. It is clear that the positive relation between these three ratios and the share price is very significant. In addition, the correlation coefficients of these relations are relatively high. The best is equal to 0.776 for the relevance between stock prices and Turnover ratio of total assets. Comparing with the results obtained from the Pearson test, the ROA, Turnover ratio of inventory, and EBIT also present a weak influence on stock prices for the Kendall test or the Spearman correlation test at the 0.05 level.

4.3. Empirical Analysis of the Steel Industry

The correlation analysis pointing to the most significant level cases can be found in Annex 3 (Tables A10-A14). Results can be divided into two parts. Only six ratios, that are the gross profit rate, ROA, EPS, BVPS, current ratio and debt-to-assets ratio passed the correlation significance tests. The BVPS and debt-to--assets ratio merely achieve the significant level of 0.01. The other four ratios, at the 0.05 significance level, show a relatively weaker relation with share prices than the former. With the same significant level, BVPS shows a positive relationship with share price; the linear dependent coefficient is equal to 0.738, while the dependent coefficient of debt-to-asset rate and closing price is −0.629. It can be seen that these six ratios belong to profitability, benefit ability to shareholders and solvency respectively. These results indicate that investors should pay more attention on these three than on others.

From the Tables (in Annex 3), it is clear that profitability, benefit ability to shareholders and solvency of public firms have a correlation with stock prices, especially the gross profit rate, the net profit rate, the ROA, the BVPS, the current ratio, and the debt-to-asset, all achieving the significant level of the 0.01 Pearson correlation test. The BVPS may be regarded as the most outstanding ratio, showing a positive correlation coefficient at 0.758. The trend of debt-to-asset ratio with prices is similar to 2011, showing a significant negative dependence. Meanwhile, the ROE, the EPS, the rate of capital accumulation and the quick ratio are tested with the significant level of 0.05.

The net profit rate, ROA, EPS, cash maturity debt ratio, debt coverage ratio, current ratio, quick ratio and debt-to-asset ratio, reach the significant level of 0.01. The highest value (0.882) is for the current ratiostock prices relationship. The Gross profit rate, rate of capital accumulation and equity ratio show a relatively lower correlation than the current ratio, achieving a significant level of 0.05. The debt-to-asset ratio and equity ratio only indicate a significant negative correlation with the share prices in 2014 exhibiting a −0.778 and −0.588 value.

Seven ratios are found to have a Pearson correlation with the share prices for all factors. Among these seven ratios, two factors, the equity ratio and the comprehensive leverage coefficient have a low relevance with a significant level of 0.05, while the other five ratios achieve the level 0.01 of significance. The highest correlation value is 0.782 between the cash maturity debt ratio and the share price. For the negative relevance side, the debt-to-asset ratio shows a bigger impact on the share price than the equity ratio.

The period from 2011 to 2015 witnesses a significant relation between 25 ratios and the stock prices through the Pearson test. The most significant numbers are for the profitability, the gross profit rate, the net profit rate and the ROA; for the benefit ability to shareholders, highest numbers are observed for the EPS and the BVPS while for the cash flow indicators, the debt coverage ratio is the highest indicator for solvency, the current ratio, the quick ratio and the debt-to-asset ratios.

In short, the EPS, the BVPS and the debt-to-asset appear to be the most significant parameters while the gross profit rate, net profit rate, ROA as well as ROE have a weak link with the investment value according to the Kendall correlation test and Spearman correlation test. In contrast, these tests indicate strong correlations between profitability, as well as benefit ability with share prices. Solvency and EBIT ratio of risk level index achieve the significant level of 0.05.

In 2014, the current ratio and debt-to-asset indicate a significant correlation with stock prices. Six factors (net profit rate, ROA, ROE, EPS, BVPS and equity ratio) have a weak impact on the share value according to the Kendall correlation test and the Spearman correlation test. The Current ratio, quick ratio and debt-to-asset have a high significant influence (0.01) to stock prices under the Kendall test as well as for the Spearman test. The equity ratio and the change of EPS can be regarded as having a relatively close relationship under the Kendall test, while the Spearman correlation test does not imply such a deduction. These two tests show a weak impact on share prices is for the ROA, cash ratio and turnover of total capital.

The net profit rate is the only factor with an important influence on the price, as it can be deduced not only from the Kendall test but also from the Spearman correlation test. At the opposite, the ROA, ROE, EPS, current ratio, debt-to-asset, equity ratio as well as EBIT show a weak relation with stock prices.

According to the Kendall test and the Spearman correlation test, the net profit rate, ROA, ROE, EPS, BVPS, current ratio, debt-to-asset ratio as well as equity ratio have a significant relationship with stock prices. In the indicators of annual reports of public steel industry, the gross profit rate, the debt security rate and the current rate exhibit a linear relationship with the stock prices while the ROE as well as the equity are the only two with one relationship—relevance to the investment value.

5. Discussion

In this section, we discuss the major statistical results observed in our empirical analysis, from an economic/financial point of view. We first examine these results for each industry; afterwards, a "cross-industrial discussion" is proposed in the second part of this section.

5.1. Media Industry

From the statistical results, Current assets turnover and the Total assets turnover for almost all most the years have a significant influence on stock prices with a high correlation coefficient. These two ratios measure the operating capacity of the public firms.

In addition, the rate of capital accumulation, a factor measuring some development ability of media enterprises, also always influences the stock prices. We can also conclude that a relationship between stock prices and influencing factors in the media industry is different from the ones in the steel industry. The current assets turnover, the total assets turnover and the rate of capital accumulations had a significant correlation with investment value period between 2011 and 2015. These assertions are in line with Shira et al. [31] who explained that, in recent years, the media industry is perceived as the major supported industry of the country.

In contrast to the steel industry and the power industry, the media industry has relatively less assets than the other two: the production equipment needed by movies and television are usually rented (such as the stage, scene, photo studio, and printing equipment). It means that the current assets turnover has a rising trend. In the same time, this tendency also accelerates the development of Total assets turnover, which leads to an increase of net profits. Next, the current assets turnover and the total asset turnover belong to measuring ratios of the operating ability. The bigger these ratios are, the stronger the ability of continuing operating will be. To ensure the ability of continuing operating, media firms pay more attention to the capability of development. A sizable amount of funds is important for firms when they try to expand their business. However, as a developing industry, it is difficult for media companies to finance themselves by issuing bonds, —when they want to realize more income. Therefore, they prefer to issue shares to acquire more funds. This situation is observable in the rate of capital accumulations. In addition to these facts, 2014 shows a surprising result in comparison to other years. There is no variable displaying a relevant correlation to stock price in 2014, but there are three correlated variables doing so in the other four years. A comment on this specificity is to be found below.

5.2. Power Industry

The results from the correlation analysis indicate that stock prices in the public power industry are more sensitive to the cash ratio, the cash maturing debt ratio, the net operating cash flow, and the total debt. These four ratios pass the Pearson test with high correlation coefficient values. These factors represent the operating capacity of the power companies. Meanwhile, the inventory turnover and the Total assets turnover, measuring operating capacity seem to have a close relationship with the stock prices. These indicators appropriately summarize the real situation of the power industry. The most significant factors measuring the solvency ability of firms, the current ratio and the quick ratio, are observed to have a strong relation with stock prices. Additionally, the ROA can also influence the stock price. In this context, the disclosed annual information shows a significant reaction upon the investment value.

Four ratios (Cash ratio, Cash maturing debt ratio, Net operating cash flow indicator and Total debt) out of 9 factors cash flow indicators, show a strong relation with the investment value. It seems that although the power industry is experiencing a mature period, an excessive expansion phenomenon occurs. The reason might be attributed to the increasing of population: the growth rate of electricity consumption in China is 14.41% per year, leading to a noticeable increase of the demand in equipment, as well as of coal and other futures. In addition to this, energy saving and environmental protection policy, as recommended by the Chinese government, imply that funds are needed to be invested into research on new energy and related technology. All these reasons result in much fund demand by power firms. Companies usually refinance their planed growth with internal funds, which can practically improve their ratios as well as decrease their finance cost. A low Net operating cash flow brings a high stock price. Apart from reinvestment from internal funds, power companies also finance their operation by issuing bonds (because of tax shields) rather than by issuing shares [32]. This is why, in the liabilities structure from power firms, there is higher percentage of long debt in the total debt. Instead of selling fixed assets to repay debt, companies merely rely on achieving more profit to meet their liabilities under the premise of continuous operation. Therefore, the lower the debt is, the higher the share price will be.

From the data analysis, it is also clear that a high current ratio and a high quick ratio, lead to a high share price. Also the ROA (return of assets) also plays a role in measuring the profitability of firms in annual reports and it influences the stock price of public power enterprises. If companies cannot achieve some income from daily operation, it will be difficult for them to repay the debt, and even to continue operating. However, a huge income not only can make power companies meet their liabilities, but also helps some reinvestment, thus with lower external financing. In addition, inventory turnover, the ratio of measuring operating capacity also shows some relevance to the share price. As one of the public sectors, the power price controlled by the government cannot cover the increasing number of coal price, which means that it may be difficult to achieve profit if the coal cost still rises [33]. In other words, a higher inventory turnover may directly lead to a stock price decrease.

5.3. Steel Industry

Our observations show that the existence of a significant correlation of stock prices with the gross profit rate, the net profit rate, the ROA (return on assets) and the ROE (return on equity). All these indicators passed the Pearson correlation test with high correlation coefficients. These ratios represent the profitability of companies. Additionally, factors measuring the benefit ability for shareholders, including EPS (earning per share) and BVPS (book value per share) are directly related to the investment decision of investors in the steel industry. The current ratio, quick ratio and debt-to-assets ratio can influence the stock price. Hence, investors need to make sure that the companies have a solvency ability (or not) before making investment decisions. At the same time, the Debt coverage ratio from the cash flow indicators also shows a significant link with stock prices. Therefore, cash flow indicators need to be considered in decision investment. In contrast to these, the ratios related to operating capacity and development ability show no real relationship with share prices.

Several trends can be observed from our analysis. First, stock prices of steel companies are more sensitive to balance sheets than to income statements. It can be seen that current ratio, quick ratio, debt-to-assets ratio, equity ratio, BVPS (book value of net assets per share), and ROA (return on assets) are closely related to assets, debts and equity of a company, which are all on the balance sheets. Actually this is not a surprising result, since steel companies need a huge amount of assets, especially properties, plants, and equipment to ensure the production and the continuity of operations. If a steel company wants to realize more income, it has two solutions: increase the price of goods, or/and expand its production capacity; the latter is leading to the need of more plants and equipment. However, more plants and equipment imply more investments; the funding may be done by issuing bonds and shares. According to corporate finance theory, debt finance has less cost than equity finance due to the existence of tax shields [34]. Hence most steel companies issue new bonds to finance new production plants and equipment; naturally the levels of total assets and debts, as well as the liquidity, become key criteria when considering investing on a steel company. For this industry, we can also notice that a higher debt-to-assets ratio brings a lower stock price; also high current ratios and quick ratios, which are the most important factors measuring the solvency ability of a company, directly lead to a higher stock price. In addition, the ROA is also related to the stock price of steel companies. This is understandable because the ROA measures the profitability of assets of a company: if a steel company cannot earn more income, it may fall into a dangerous condition because of facing the difficulty to repay huge debts. In the same vein, gross profit rate and net profit rate of a steel company are essential to support its stock price. If a company is able to earn enough money to invest on more plants and equipment, it may not need to finance much from debts, thereby lowering its operating risk and giving investors a clearer and more optimistic outlook.

Interestingly, the correlated variables in 2015 are a bit different from the variables between 2011 and 2014. There are only 4 variables correlated with stock prices in 2015, while there are at least 6 correlated variables in the other 4 years. Gross profit rate, net profit rate, ROA, and EPS are not among the correlated variables. There is a specific reason why this abnormal result appears: due to the lower economic growth rate and total steel demand of China, steel industry experienced a recession in 2015; consequently steel companies did not make much profit in 2015 and many of them faced losses. Such a situation explains why no predictive conclusion can be actually drawn for the next year(s) from such values. In order to stimulate the steel industry, the Chinese government issued some policies in early 2016, which greatly strengthened the stock price trend of steel companies. However, companies in China revealed their annual reports in March and April of 2016, when the market sentiment was totally optimistic towards steel industry. Therefore, the losses mentioned on annual reports of steel companies were not able to affect the stock price. This phenomenon contributes much to the abnormal result found in our study.

5.4. Cross-Industrial Discussion

In view of the above statistical results, it can be claimed that there are many differences about the correlations between ratios derived from published annual reports and stock prices for the three industries. In the media industry, the development ability is proved to have a significant relation with the stock prices, while there is no clear influence on the share price in the power industry and in the steel industry. The media industry is experiencing the sunrise period of total industry lifecycle with well accretive feature. It means that the media industry has a unique pattern in transporting information and achieving profits. The main business pattern of media firms is not to be directly selling media product, but enlarging its audience and the range of media transmission, as far and as widely as possible, to build a profit. This structure can be particularly found in media sub-sectors, such as internet, wireless televisions, and newspapers. Because of this, the feature of the media industry includes lower costs, broader impacts and stronger profitability as compared with other industries. More funds flow into the media industry rather than to other industrial sectors, which leads to higher expected profits in contrast to the cases in other sectors. Media firms also face with a fiercer competition.

Therefore, each company tries its best to expand the scope of the enterprise, such as building new projects, using its internal funds or financing by issuing bonds as well as shares, rather than sharing out dividends. In this moving context, there is no regular pattern in the evolution of fundamentals and investors mainly focus on ratios measuring the operating ability as well as development capacity of public companies. Furthermore, 2014 appears to be a specific year with a new regulation on media industry in China where the sociopolitical context influenced the industry much more than financial results [35]. That could be a reason for why there is no significant relationship between variables and stock price of the media industry in 2015.

For the power and steel industry, the ROA, the current ratio and the quick ratio (the most important factor on solvency) showed a significant, relevance for investment decisions. However, the significant level and the correlation coefficient of these relations in these two fields are different. In general, ratios from profitability and solvency for public steel firms will predict in a more accurate way the tendency of the stock price rather than those in the power industry.

The power industry has reached a mature cycle-life. This is relatively stable and clear industry. At this time, the profitability of the power industry is decreasing, along with more difficulties to research new products. Nevertheless, the improvement of the power industry has a high relevance to other industries, such as the coal industry, the petroleum industry, the transportation industry etc. Because this industry plays a significant role in the Chinese economy, the government introduced some supportive policies in the power field that guarantees a relatively stable tendency of profit margin. The main source of income for power enterprises is when selling electricity to residents as well as companies with cash settlement rather than in debt of receivables. Therefore, more cash in the financial reports means higher sales volume of electricity, directly leading to a better performance measure. Thus, confronting a relatively stable power industry, investors tend to follow a fundamental analysis approach rather than a technical one. We propose that it is the reason why the cash flow indicators themselves show a close relation to stock price in the above tests on the power industry.

Regarding the Chinese steel industry, the production capacity of steel companies exceeds market demand. In this situation of oversupply, although there is an increase trend of raw materials, the steel price cannot rise. This condition means that it is difficult to change the status of falling profits in the steel field. Production and sales of automobiles shows a slight growth recently [36]. In general, the prospects of downstream industries, which need steel during the operating activities, are not satisfactory. A lower demand of steel from this market directly leads to a lower profitability of the steel firms. Thus, the state of operations of the steel industry is not optimistic, along with a lower profitability with respect to that of other industries. However, because there is no aim for expanding companies and developing new projects, steel enterprises, share their profit as dividends. Investors, who put their funds into public steel companies, also pursue dividends rather than arbitrage on a short term. In this respect, investors prefer to study financial ratios from annual reports in order to measure in which public companies it is worth to invest. Therefore, stock prices in the steel field can be in a much closer relationship to the financial information than a priori expected.

6. Conclusions

This paper analyzes the correlations of financial ratios from published annual reports with the stock prices in the media industry, the power industry as well as the steel industry in China. These correlations are measured by the Pearson test, the Kendall test and the Spearman correlation test. In a transitional environment as the one implemented in China, the quality of information and decision tools are often questioned [7]. Our study emphasizes the importance of fundamental analysis as a predictive tool in China. Precisely, a statistical study of the relationship between financial ratios and stock prices illustrates quite well the diversity and the heterogeneity of the economic context in China, since, depending on the sector under consideration, diverse ratios have different impact on the stock prices :
(1) The stock price of media companies only has a relation to the financial factors which measure their development ability and their operating capacity.
(2) Investors, buying stocks from public power firms, must pay attention on cash flow indicators. Apart from this, factors about profitability, operating ability and solvency capacity are also closely related to investment values.
(3) Different from the former two industries, factors measuring the operating capacity and development ability from the steel annual reports exhibit an influence on the share prices.

This article only focused on the descriptive correlation analysis emphasizing the statistical relationship between fundamental analysis and stock prices, showing therefore that the former can effectively be used as an investment tool. The objective of this descriptive study was to investigate the economic meaning of the fundamental analysis. In contradiction to the EMH for which the accounting information is instantaneously integrated into the financial information (stock prices), our results suggest that these two levels of information might not be synchronized opening therefore a door for a fundamental analysis based prediction. Our empirical study supports this perspective. Furthermore, our results also indicate that accounting information illustrate quite well the economic reality since financial reports in each industry can disclose a part of stock value information in line with the economic situation of the industry under consideration. To some extent, our study can be seen as a first step of for further econometric investigations on the statistical link between account indicators and stocks prices.

Conflicts of Interest The authors declare no conflicts of interest regarding the publication of this paper.

| Ratio | Calculation |
|---|---|
| Gross profit rate | (Sales profit/Sales income)×100 |
| Net profit rate | (Net profit/Sales income)×100 |
| Return on Assets (ROA) | (Net profit/Average assets in accounting)×100 |
| Return on Equity (ROE) | (Net profit/Average equity in accounting)×100 |
| Earnings per Share (EPS) | Net earnings/Number of shares outstanding |
| Book value per Share (BVPS) | Book value of net asset/Number of shares outstanding |

Table 1. Six Profitability Ratios

| Ratio | Calculation |
|---|---|
| Cash ratio | (Cash+Securities)×100 /Current liabilities |
| Cash maturing debt ratio | Net Operating Cash Flow/Maturing debt |
| Debt coverage ratio | Net Operating Cash Flow/Net Profit |
| Net cash flow per share | Net cash flow/Number of shares outstanding |

Table 2. Four Cash Flow Indicators

| Ratio | Calculation |
|---|---|
| Turnover of account receivable | Sales income/Average account receivable |
| Inventory turnover | Sales cost /Average inventory |
| Current assets turnover | Sales income/Average current assets |
| Total assets turnover | Sales income/Average assets |

Table 3. Four Operating Capacity Indicators

| Ratio | Calculation |
|---|---|
| Capital accumulation rate | Increase of equity/Equity at beginning of period |
| Growth rate of EPS | Increase of EPS/EPS at beginning of period |
| Growth rate of ROE | Increase of ROE/ROE at beginning of period |
| Net Profit growth rate | Increase of net Profit/net profit at beginning of period |

Table 4. Four Development Ability Ratios

| Ratio | Calculation |
|---|---|
| Current ratio | Current assets/Current liabilities |
| Quick ratio | (Current assets - inventories) / Current liabilities assets- |
| Debt-to-assets ratio | Liabilities/Assets |
| Equity ratio | Liabilities/Shareholders' Equity |
| Financial leverage | Change rate of EPS/Change rate of EBIT |
| Operating leverage | Change rate of EBIT/Change rate of sales amount |
| Degree of total leverage | Financial leverage × Operating leverage |

Table 5. Four Solvency Ability and three Risk Level Ratios

Annex 1

The correlation test results between financial ratios and stock prices for the media industry in years 2011-2015; **Correlation is significant at the 0.01 level (2tailed), reddened cells; *Correlation is significant at the 0.05 level (2tailed), yellowed cells.

(Table 6) Table A1. The correlation test results between financial ratios and stock prices for the media industry in 2011

| 2011 (N=8) | Current assets turnover | Total assets turnover | |
|---|---|---|---|
| Pearson Correlation | 0.809* | 0.754* | |
| Sig. | 0.015 | 0.031 | |
| | Degree of operating leverage | Total sale change | Comprehensive leverage coefficient |
| Kendall Correlation | -0.571* | 0.571* | -0.643* |
| Sig. | 0.048 | 0.048 | 0.026 |
| Spearman Correlation | -0.738* | 0.786* | -0.786* |
| Sig. | 0.037 | 0.021 | 0.021 |

(Table 7) Table A2. The correlation test results between financial ratios and stock prices for the media industry in 2012

| 2012 ( N = 8 ) | Cash ratio | Current assets turnover | Total assets turnover | Current ratio |
|---|---|---|---|---|
| Pearson Correlation | -0.799* | 0.932** | 0.709* | -0.746* |
| Sig. | 0.017 | 0.001 | 0.049 | 0.033 |
| | Quick ratio | Debt-to-assets ratio | Equity ratio | |
| Pearson Correlation | -0.735* | 0.789* | 0.819* | |
| Sig. | 0.038 | 0.02 | 0.013 | |
| | Cash ratio | Current assets turnover | Current ratio | |
| Kendall Correlation | -0.857** | 0.857** | -1.000** | |
| Sig. | 0.003 | 0.003 | | |
| Kendall Correlation | Quick ratio | Debt-to-assets ratio | Equity ratio | |
| Sig. | -0.929** | 0.571* | 0.714* | |
| | 0.001 | 0.048 | 0.013 | |
| | Cash ratio | Current assets turnover | Current ratio | |
| Spearman Correlation Sig. | -0.929** | 0.929** | -1.000** | |
| | 0.001 | 0.001 | | |
| | Quick ratio | Debt-to-assets ratio | Equity ratio | |
| Spearman Correlation Sig. | -0.976** | 0.714* | 0.810* | |
| | 0.000 | 0.047 | 0.015 | |

(Table 8) Table A3. The correlation test results between financial ratios and stock prices for the media industry in 2013.

| 2013 ( N = 8 ) | Capital accumulation rate |
|---|---|
| Pearson Correlation Sig. | 0.809* |
| | 0.015 |
| | Inventory turnover |
| Spearman Correlation Sig. | 0.738* |
| | 0.037 |

(Table 9) Table A4.. The correlation test results between financial ratios and stock prices for the media industry in 2015.

| 2015 ( N = 17 ) | Capital accumulation rate | Growth rate of EPS | Growth rate of ROE | Net profit growth rate |
|---|---|---|---|---|
| Pearson Correlation Sig. | 0.573* | 0.625** | 0.588* | 0.644** |
| | 0.016 | 0.007 | 0.013 | 0.005 |
| Kendall Correlation Sig. | BVPS | Capital accumulation rate | Degree of operating leverage | |
| | 0.391* | 0.450* | 0.391* | |

|  | BVPS | Rate of capital accumulation | Degree of operating leverage |  |
|---|---|---|---|---|
|  | 0.029 | 0.012 | 0.029 |  |
| Spearman Correlation Sig. | 0.591* | 0.655** | 0.478* |  |
|  | 0.012 | 0.004 | 0.052 |  |

**Annex 2** The correlation test results between financial ratios and stock prices for the power industry in years 2011-2015; **Correlation is significant at the 0.01 level (2-tailed), reddened cells; *Correlation is significant at the 0.05 level (2-tailed), yellowed cells.

(Table 10) Table A5. The correlation test results between of financial ratio and stock prices in various years for the power industry; the number of relevant companies N is given for the corresponding year

| 2011 ( N = 18) | ROA | ROE | EPS | Net cash flow per share |  |
|---|---|---|---|---|---|
| Pearson Correlation Sig. | 0.497* | 0.727** | 0.839** | -0.615** |  |
|  | 0.036 | 0.001 | 0.000 | 0.007 |  |
|  | Inventory turnover | Current assets turnover | Total assets turnover |  |  |
| Pearson Correlation Sig. | -0.552* | -0.493* | 0.506* |  |  |
|  | 0.018 | 0.038 | 0.032 |  |  |
|  | ROA | ROE | EPS | Inventory turnover | Equity ratio |
| Kendall Correlation Sig. | 0.354* | 0.367* | 0.383* | -0.459** | -0.375* |
|  | 0.041 | 0.034 | 0.028 | 0.008 | 0.031 |
|  | ROA | ROE | EPS | Inventory turnover | Equity ratio |
| Spearman Correlation Sig. | 0.511* | 0.491* | 0.503* | -0.639 | -0.513 |
|  | 0.030 | 0.038 | 0.033 | 0.004 | 0.029 |
|  | Total assets turnover | EPS growth rate | Current ratio | Quick ratio |  |
| Kendall Correlation Sig. | 0.502** | 0.354* | 0.459** | 0.380* |  |

|  | 0.004 | 0.041 | 0.008 | 0.028 |  |
| --- | --- | --- | --- | --- | --- |
|  | Total assets turnover | EPS growth rate | Current ratio | Quick ratio |  |
| Spearman Correlation | 0.707 | 0.542 | 0.605 | 0.491 |  |
| Sig. | 0.001 | 0.02 | 0.008 | 0.038 |  |

(Table 11) Table A6. The correlation test results between financial ratios and stock prices for the power industry in 2012

| 2012 (N = 18) | ROA | ROE | EPS | Cash ratio |
| --- | --- | --- | --- | --- |
| Pearson Correlation | 0.591** | 0.638** | 0.779** | 0.480* |
| Sig. (2-tailed) | 0.010 | 0.004 | 0.000 | 0.044 |
|  | Cash maturing debt ratio | Net cash flow per share | Current ratio |  |
| Pearson Correlation | 0.489* | -0.631** | 0.570* |  |
| Sig. (2-tailed) | 0.039 | 0.005 | 0.014 |  |
|  | Current assets turnover | Current ratio | Operating net cash flow |  |
| Kendall Correlation | 0.503** | 0.386* | -0.386* |  |
| Sig. (2-tailed) | 0.004 | 0.025 | 0.025 |  |
|  | Operating net cash flow | Current assets turnover | Current ratio | Degree of total leverage |
| Spearman Correlation | -0.560* | 0.713** | 0.501* | -0.474* |
| Sig. (2-tailed) | 0.016 | 0.001 | 0.034 | 0.047 |

(Table 12) Table A7. The correlation test results between financial ratios and stock prices for the power industry in 2013.

| 2013 (N = 19) | ROA | EPS | Cash ratio | Operating net cash flow | Inventory turnover |
| --- | --- | --- | --- | --- | --- |
| Pearson Correlation | 0.573* | 0.457* | 0.589** | -0.551* | -0.463* |

| | Sig. | 0.010 | 0.049 | 0.008 | 0.014 | 0.046 |
|---|---|---|---|---|---|---|
| Pearson Correlation Sig. | | Total assets turnover | Current ratio | Quick ratio | EBIT | |
| | | -0.475* | 0.744** | 0.738** | -0.510* | |
| | | 0.040 | 0.000 | 0.000 | 0.026 | |
| Kendall Correlation Sig. | | Cash ratio | Operating net cash flow | Inventory turnover | Current ratio | Quick ratio |
| | | 0.413* | -0.427* | -0.368* | 0.516** | 0.516** |
| | | 0.014 | 0.011 | 0.028 | 0.002 | 0.002 |
| Spearman Correlation Sig. | | Cash ratio | Operating net cash flow | Inventory turnover | Current ratio | Quick ratio |
| | | 0.511* | -0.589** | -0.514* | 0.638** | 0.618** |
| | | 0.025 | 0.008 | 0.024 | 0.003 | 0.005 |
| Spearman Correlation Sig. | | Current assets turnover | Total assets turnover | | | |
| | | -0.482* | -0.478* | | | |
| | | 0.036 | 0.038 | | | |

(Table 13) Table A8. The correlation test results between financial ratios and stock prices for the power industry in 2014.

| 2014 (N = 20) | Cash maturing debt ratio | Operating net cash flow | Inventory turnover | Current assets turnover |
|---|---|---|---|---|
| Pearson Correlation Sig. (2-tailed) | 0.537* | -0.581** | -0.573** | -0.493* |
| | 0.015 | 0.007 | 0.008 | 0.027 |
| | Current ratio | Quick ratio | EBIT | Total assets turnover |
| Pearson Correlation Sig. (2-tailed) | 0.513* | 0.456* | -0.561* | 0.692** |
| | 0.021 | 0.043 | 0.010 | 0.001 |
| | Cash ratio | Operating net cash flow | Inventory turnover | Total assets turnover |
| Kendall Correlation Sig. (2-tailed) | 0.324* | -0.358* | -0.347* | 0.491** |

|  | 0.047 | 0.027 | 0.032 | 0.003 |
|---|---|---|---|---|
| Spearman Correlation Sig. (2-tailed) | Cash ratio | Operating net cash flow | Inventory turnover | Total assets turnover |
|  | 0.452* | -0.520* | -0.507* | 0.692** |
|  | 0.046 | 0.019 | 0.023 | 0.001 |
| Kendall Correlation Sig. (2-tailed) | Current ratio |  | EBIT | Quick ratio |
|  | 0.442** |  | -0.368* | 0.421** |
|  | 0.006 |  | 0.023 | 0.009 |
| Spearman Correlation Sig. (2-tailed) | Current ratio |  | EBIT | Quick ratio |
|  | 0.571** |  | -0.525* | 0.550* |
|  | 0.008 |  | 0.018 | 0.012 |

(Table 14) Table A9. The correlation test results between financial ratios and stock prices for the power industry in 2015

| 2015 (N = 20) | Cash maturing debt ratio | Operating net cash flow | Capital accumulation rate | Total assets turnover | Cash ratio |
|---|---|---|---|---|---|
| Pearson Correlation Sig. | 0.539* | -0.487* | 0.540* | 0.776** | 0.493* |
|  | 0.014 | 0.029 | 0.014 | 0.000 | 0.027 |
|  | Current ratio | Quick ratio | Net profit growth rate | Degree of operating leverage | Degree of total leverage |
| Pearson Correlation Sig. | 0.608** | 0.524* | 0.513* | 0.540* | 0.572** |
|  | 0.004 | 0.018 | 0.021 | 0.014 | 0.008 |
|  | ROA | Cash ratio | Operating net cash flow | Inventory turnover | Total assets turnover |
| Kendall Correlation Sig. | 0.368* | 0.421** | -0.484** | -0.337* | 0.575** |
|  | 0.023 | 0.009 | 0.003 | 0.038 | 0 |
| Spearman Correlation Sig. | ROA | Cash ratio | Operating net cash flow | Inventory turnover | Total assets turnover |
|  | 0.496* | 0.538* | -0.677** | -0.481* | 0.749** |

|  | 0.026 | 0.014 | 0.001 | 0.032 | 0.000 |
|---|---|---|---|---|---|
|  | Current ratio | Quick ratio | EBIT | Degree of total leverage |  |
| Kendall Correlation Sig. | 0.501** | 0.512** | -0.389* | 0.337* |  |
|  | 0.002 | 0.002 | 0.016 | 0.038 |  |
|  | Current ratio | Quick ratio | EBIT | Degree of total leverage |  |
| Spearman Correlation Sig. | 0.662** | 0.663** | -0.561* | 0.481* |  |
|  | 0.001 | 0.001 | 0.010 | 0.032 |  |

## Annex 3

The correlation test results between financial ratios and stock prices for the steel industry in [2011-2015].

(Table 15) Table A10. The correlation test results between financial ratios and stock prices for for the steel industry in 2011

| 2011 (N = 17) | Gross profit rate | ROA | EPS | BVPS | Current ratio | Debt-to-assets ratio |
|---|---|---|---|---|---|---|
| Pearson Correlation Sig. | 0.529* | 0.498* | 0.569* | 0.685** | 0.545* | -0.629** |
|  | 0.029 | 0.042 | 0.017 | 0.002 | 0.024 | 0.007 |

|  | Gross profit rate | Net profit rate | ROA | ROE |
|---|---|---|---|---|
| Kendall Correlation Sig. | 0.332 | 0.391* | 0.421* | 0.303 |
|  | 0.064 | 0.029 | 0.019 | 0.091 |
| Spearman Correlation Sig. | 0.490* | 0.536* | 0.587* | 0.499* |
|  | 0.046 | 0.027 | 0.013 | 0.041 |
|  | EPS | BVPS | Debt-to-assets ratio |  |
| Kendall Correlation Sig. | 0.607** | 0.533** | -0.524** | |
| | 0.001 | 0.003 | 0.003 | |
| Spearman Correlation Sig. | 0.712** | 0.710** | -0.684** | |
| | 0.001 | 0.001 | 0.002 | |

(Table 16) Table A11. The correlation test results between financial ratios and stock prices for the steel industry in 2012.

| 2012 (N = 19) | Gross profit rate | Net profit rate | ROA | ROE | EPS |
|---|---|---|---|---|---|
| Pearson Correlation Sig. | 0.616** | 0.693** | 0.675** | 0.476* | 0.514* |
| | 0.005 | 0.001 | 0.002 | 0.039 | 0.024 |
| | BVPS | Rate of capital accumulation | Current ratio | Quick ratio | Debt-to-assets ratio |
| Pearson Correlation Sig. | 0.758** | 0.566* | 0.616** | 0.506* | -0.578** |
| | 0.000 | 0.011 | 0.005 | 0.027 | 0.010 |

| | Net profit rate | ROA | ROE | EPS | |
|---|---|---|---|---|---|
| Kendall Correlation Sig. | 0.474** | 0.556** | 0.462** | 0.448** | |
| | 0.005 | 0.001 | 0.006 | 0.008 | |
| Spearman Correlation Sig | 0.582** | 0.679** | 0.605** | 0.562* | |
| | 0.009 | 0.001 | 0.006 | 0.012 | |
| | BVPS | Rate of capital accumulation | EPS Growth rate | ROE Growth rate | |
| Kendall Correlation Sig. | 0.544** | 0.439** | 0.532** | 0.509** | |
| | 0.001 | 0.009 | 0.001 | 0.002 | |
| Spearman Correlation Sig. | 0.718** | 0.612** | 0.705** | 0.675** | |
| | 0.001 | 0.005 | 0.001 | 0.002 | |
| | Net profit growth rate | Current ratio | Quick ratio | Debt-to-assets ratio | EBIT |
| Kendall Correlation Sig. | 0.556** | 0.406* | 0.352* | -0.345* | 0.333* |
| | 0.001 | 0.016 | 0.036 | 0.039 | 0.046 |
| Spearman Correlation Sig. | 0.691** | 0.553* | 0.480* | -0.498* | 0.437* |
| | 0.001 | 0.014 | 0.038 | 0.030 | 0.061 |

(Table 17) Table A12. The correlation test results between financial ratios and stock prices for the steel industry in 2013

| 2013 (N = 17) | Gross profit rate | Net profit rate | ROA | EPS | BVPS |
|---|---|---|---|---|---|
| Pearson Correlation Sig. | 0.785** | 0.483* | 0.647** | 0.505* | 0.585* |
| | 0.000 | 0.050 | 0.005 | 0.039 | 0.014 |
| | Debt coverage ratio | Current ratio | Quick ratio | Debt-to-assets ratio | Equity ratio |
| Pearson Correlation Sig. | 0.639** | 0.829** | 0.665** | -0.747** | -0.550* |
| | 0.006 | 0.000 | 0.004 | 0.001 | 0.022 |

| | Net profit growth rate | ROA | ROE | EPS | BVPS |
|---|---|---|---|---|---|
| | 0.421* | 0.391* | 0.406* | 0.442* | 0.406* |
| Kendall Correlation Sig. | 0.019 | 0.029 | 0.023 | 0.015 | 0.023 |
| | Net profit growth rate | ROA | ROE | EPS | BVPS |
| | 0.546* | 0.540* | 0.538* | 0.565* | 0.553* |
| Spearman Correlation Sig. | 0.023 | 0.025 | 0.026 | 0.018 | 0.021 |
| | Current ratio | Quick ratio | Debt-to-assets ratio | Equity ratio | EPS |
| Kendall Correlation Sig. | 0.509** | 0.352 | -0.568** | -0.415* | 0.442* |
| | 0.004 | 0.052 | 0.002 | 0.021 | 0.015 |
| | Current ratio | Quick ratio | Debt-to-assets ratio | Equity ratio | EPS |
| Spearman Correlation Sig. | 0.667** | 0.538* | -0.715** | -0.523* | 0.565* |

|  |  | 0.003 | 0.026 | 0.001 | 0.031 | 0.018 |

(Table 18) Table A13. The -correlation test results between financial ratios and stock prices for the steel industry in 2014.

| 2014 (N=18) | Gross profit rate | Net profit rate | ROA | EPS | Cash maturing debt ratio | Debt coverage ratio |
|---|---|---|---|---|---|---|
| Pearson Correlation | 0.562* | 0.672** | 0.737** | 0.601** | 0.734** | 0.599** |
| Sig. | 0.015 | 0.002 | 0.000 | 0.008 | 0.001 | 0.009 |
|  | Rate of capital accumulation | Current ratio | Quick ratio | Debt-to-assets ratio | Equity ratio |  |
| Pearson Correlation | 0.480* | 0.882** | 0.744** | -0.778** | -0.588* |  |
| Sig. | 0.044 | 0.000 | 0.000 | 0.000 | 0.010 |  |
|  |  |  | ROA | Cash ratio | Total assets turnover | Current ratio |
| Kendall Correlation |  |  | 0.438* | 0.370* | 0.359* | 0.511** |
| Sig. |  |  | 0.011 | 0.034 | 0.037 | 0.003 |
|  |  |  | ROA | Cash ratio | Total assets turnover | Current ratio |
| Spearman Correlation |  |  | 0.579* | 0.501* | 0.521* | 0.696** |
| Sig. |  |  | 0.012 | 0.034 | 0.027 | 0.001 |
|  |  |  | Quick ratio | Debt-to-assets ratio | Equity ratio |  |
| Kendall Correlation |  |  | 0.454** | -0.503** | -0.446** |  |
| Sig. |  |  | 0.009 | 0.004 | 0.010 |  |
|  |  |  | Quick ratio | Debt-to-assets ratio | Equity ratio |  |
| Spearman Correlation |  |  | 0.631** | -0.604** | -0.576* |  |
| Sig. |  |  | 0.005 | 0.008 | 0.012 |  |

(Table 19) Table A14. The correlation test results between financial ratios and stock prices for the steel industry in 2015.

| 2015 (N = 17) | Cash maturing debt ratio | Debt coverage ratio | Current ratio | Quick ratio | Debt-to-assets ratio | Equity ratio | Degree of total leverage |
|---|---|---|---|---|---|---|---|
| Pearson Correlation | 0.782** | 0.648** | 0.756** | 0.634** | -0.765** | -0.491* | 0.528* |
| Sig. | 0.000 | 0.005 | 0.000 | 0.006 | 0.000 | 0.045 | 0.029 |

|  | Net profit growth rate | ROA | ROE | EPS | Current ratio |
|---|---|---|---|---|---|
| Kendall Correlation | 0.465** | 0.391* | 0.450* | 0.415* | 0.415* |
| Sig. | 0.009 | 0.029 | 0.012 | .021 | 0.021 |

|  | Net profit growth rate | ROA | ROE | EPS | Current ratio |
|---|---|---|---|---|---|
| Spearman Correlation | 0.629** | 0.506* | 0.585* | 0.515* | 0.503* |
| Sig. | 0.007 | 0.038 | .014 | 0.034 | 0.040 |

|  | Debt-to-assets ratio | Equity ratio | EPS | EBIT |
|---|---|---|---|---|
| Kendall Correlation | -0.406* | - 0.385* | 0.415* | 0.376* |
| Sig. | 0.023 | 0.032 | 0.021 | 0.035 |

|  | Debt-to-assets ratio | Equity ratio | EPS | EBIT |
|---|---|---|---|---|
| Spearman Correlation | -0.519* | -0.508* | 0.515* | 0.546* |
| Sig. | 0.033 | 0.037 | 0.034 | 0.023 |

**Correlation is significant at the 0.01 level (2-tailed).
*Correlation is significant at the 0.05 level (2-tailed)